\documentclass[aps,prd,preprint,superscriptaddress,showpacs,floatfix,nobibnotes]{revtex4-1}
\usepackage{hyperref}
\usepackage{comment}
\usepackage{latexsym}
\usepackage{setspace}
\usepackage{amsmath}
\usepackage{amssymb}
\usepackage{graphicx}
\usepackage{float}
\usepackage[usenames, dvipsnames]{color}
\usepackage{epsfig}
\usepackage{subcaption}
\usepackage[normalem]{ulem}

\newcommand{\E}{{\mathcal E}}

\usepackage[english]{babel} 
\usepackage{amsmath,amsfonts,amsthm} 
\usepackage{graphicx} 
\usepackage{subcaption}

\newcommand{\bea}{\begin{eqnarray}}
\newcommand{\eea}{\end{eqnarray}}

\def\threevector#1#2#3{ \left(
\begin{tabular}{c}
$#1$\\
$#2$\\
$#3$
\end{tabular}
\right)}

\begin{document}

\title{Nonequilibrium approach to holographic superconductors using gradient flow}

\author{P. Mikula}
\email[]{pnmikula@gmail.com} \affiliation{Department of Physics, Brandon University, Brandon, Manitoba R7A 6A9, Canada}
\affiliation{Department of Physics, University of Winnipeg, Winnipeg, Manitoba R3B 2E9, Canada}
\affiliation{Department of Physics and Astronomy, University of Manitoba, Winnipeg, Manitoba R3T 2N2, Canada}
\affiliation{Winnipeg Institute for Theoretical Physics, Winnipeg, Manitoba, Canada}

\author{M.E. Carrington}
\email[]{carrington@brandonu.ca} 
\affiliation{Department of Physics, Brandon University, Brandon, Manitoba R7A 6A9, Canada}
\affiliation{Winnipeg Institute for Theoretical Physics, Winnipeg, Manitoba, Canada}

\author{G.~Kunstatter}
\email[]{gkunstatter@uwinnipeg.ca} 
\affiliation{Department of Physics, University of Winnipeg, Winnipeg, Manitoba R3B 2E9, Canada}
\affiliation{Winnipeg Institute for Theoretical Physics, Winnipeg, Manitoba, Canada}

\date{\today}

\begin{abstract}

We study a charged scalar field in a bulk 3+1-dimensional anti-de Sitter (AdS) spacetime with a planar black hole background metric. 
Through the AdS/CFT correspondence this is equivalent to a strongly coupled field theory in 2+1 dimensions describing a superconductor. 
We use the gradient flow method and solve the flow equations numerically between two fixed points: a vacuum solution and a hairy black hole solution. 
We study the corresponding flow on the boundary between a normal metal phase and a superconducting phase. We show how the gradient flow moves fields between two fixed points in a way that minimizes the free energy of the system.
At the fixed points of the flow
the AdS/CFT correspondence  provides an equivalence between the Euclidean on-shell action in the bulk and the free energy of the boundary, but it does not tell us about fields away from equilibrium.
However, we can formally link static off-shell configurations in the bulk and in the boundary at the same point along the flow. 
For quasistatic evolution at least, it may be  reasonable to think of this link as an extension of the AdS/CFT correspondence. 

\end{abstract}


\normalsize
\maketitle

\tableofcontents

\normalsize

\section{Introduction}


The AdS/CFT correspondence provides a possible mechanism for studying strongly interacting quantum field theories in $d$ dimensions via the analysis of dual, weakly coupled gravitational systems in $d+1$ dimensions. 
First proposed by Maldecena ~\cite{Maldecena} in the context of string theory,
 in recent years the AdS/CFT correspondence has proven useful for studying a variety of condensed matter systems (see ~\cite{Hartnoll2012, horowitz2} for excellent reviews). One of the greatest successes of this holographic approach has been its application to the study of  quark gluon plasma~\cite{QGPlasmaReview}.    

 In this paper we are specifically interested in a gravitational theory in $3+1$ dimensions that gives a holographically dual description of a $2+1$-dimensional superconductor. The model for a holographic superconductor  that we consider was originally developed in ~\cite{gubser,hartnoll1}. The bulk theory consists of an electromagnetic field and charged scalar minimally coupled to Einstein gravity with a negative cosmological constant. Since the gravity theory is weakly coupled the superconducting theory on the boundary is strongly coupled. Useful reviews are found in Refs.~\cite{hartnoll2,horowitz1,musso}. 

Here we apply a gradient flow method to study the static solutions and nonequilibrium behavior of the holographic superconductor. Gradient flow is a general analogue of the heat equation that describes the path of steepest descent with respect to the free energy or Euclidean action of a system. It has been used in a variety of mathematical and physical contexts including Perelman's proof~\cite{Perelman2002}  of Thurston's geometrization conjecture,  string theory~\cite{StringyFlow}, superconductors~\cite{SuperFlow} and even image processing~\cite{ImageFlow}, 
\cite{FootnoteAndrei}.

First, we use gradient flow to reproduce a number of known solutions for a charged scalar field in a fixed anti-de Sitter (AdS) black hole background, and to obtain some solutions that are new to the best of our knowledge. Through the AdS/CFT correspondence, the solutions we obtain for the bulk theory
allow us to  study the properties of the strongly coupled superconductor on the boundary. We determine the critical temperature and magnetic fields below which the material becomes superconducting, and the coherence lengths for the condensate operator and the charge density.  In addition to being a numerical tool for finding solutions, the gradient flow itself provides a good test of the nonlocal, as opposed to local, stability of these solutions.  We verify this by calculating the (off-shell) Euclidean action ($S_E$) in the bulk along the flow and show that, as expected, the solutions minimize the free energy.

Secondly, we use the gradient flow to study the approach to thermal equilibrium of a holographic superconductor. Although the gradient flow is not equivalent to, or a replacement for, full time-dependent solutions, it may provide predictions for the approach to thermal equilibrium for holographic superconductors in cases where evolution is quasistatic and dissipation requires the system to be nonisolated from the environment so that energy is not conserved.
Our flow calculations are carried out in the bulk, but they create a corresponding flow in the boundary operators. By solving for the flow parameter in terms of the energy, we obtain the path that describes the progression of the boundary operators as a function of the decreasing energy. 

The paper is organized as follows:
In Sec. II we briefly introduce the holographic superconducting model that we consider and provide a brief general review of gradient flow. Section III derives  the flow equations that we solve, while Sec. IV presents the numerical methods and results. We close in Sec. V with conclusions and prospects for future work.

\section{Preliminaries}
\subsection{The holographic superconductor}
Following~\cite{hartnoll1}, we start with a planar Schwarzschild anti-de Sitter black hole
\bea
\label{metric}
ds^2 = -f(r) dt^2+\frac{dr^2}{f(r)} + r^2 d\Omega^2_{R^2} \text{~~ with~~} f(r) = \frac{r^2}{L^2} -\frac{M}{r}
\eea
where $L$ is the AdS radius, $M$ is related to the Hawking temperature of the black hole $T = 3M^{1/3}/(4\pi L^{4/3})$ and $d\Omega^2_{R^2}$ is the metric on flat two-dimensional space. 
The black hole is 3+1 dimensional and is dual to a 2+1-dimensional theory on the boundary. 
We consider $d\Omega^2_{R^2}$ in both polar and Cartesian coordinates.

For the gravitational part of the action we consider the standard Einstein-Hilbert action with a negative cosmological constant, so that \eqref{metric} is the unique (uncharged) vacuum solution. We want the  boundary theory to describe a superconductor. This requires the boundary theory to contain charged fields whose condensation can lead to superconductivity. 
A conventional $s$-wave superconductor has an isotropic condensate described by a charged scalar field. The tensor properties of the corresponding bulk operator must be the same, and therefore we  introduce a charged scalar in the bulk theory. 
We take this field to be minimally coupled to a bulk electromagnetic field. Our bulk action is therefore 
\begin{equation}
\label{action}
S[A,\psi]=\int dx^4 \sqrt{-g} \left[\frac{c^4}{16\pi G} \left(R -2\Lambda \right) -\frac{1}{4}F^{\mu\nu}F_{\mu\nu} -(D^{\mu}\psi)^{\dagger}D_{\mu}\psi  - m^2\psi^{\dagger}\psi \right],
\end{equation}
where $F_{\mu\nu} = \partial_\mu A_\nu - \partial_\nu A_\mu$, $D_{\mu} = \partial_{\mu} - i q A_{\mu}$, $\psi$ is a complex scalar field, and $\Lambda$ is the cosmological constant.

A condensate in the bulk theory would correspond to a static nonzero scalar field outside the black hole horizon, in apparent contradiction to black hole no hair theorems. These theorems are based on the idea that matter outside a black hole wants to either fall into the black hole, or radiate out to infinity in the asymptotically flat case. 
The proof is formulated in terms of a black hole uniqueness theorem, which says that when gravity is coupled to matter fields, a stationary black hole solution is uniquely characterized by its conserved charges.
However, there is no completely general no hair theorem, and counterexamples have been known since the 1990s \cite{NoHair}.

In our problem, the formation of scalar hair is possible because we work in AdS space, where the negative cosmological constant acts like a confining box that prevents the charged particles from escaping to infinity.
It is easy to see why the vacuum in the theory defined in \eqref{action} might be unstable to the formation of scalar hair. 
The effective mass of the scalar field is $m_{\rm eff}^2 = m^2 + g^{tt}q^2A_t^2 + \cdots $. Since $g_{tt} = -f(r) <0$ it is possible that the effective mass becomes sufficiently negative near the horizon to destabilize the scalar field. 

We can also see that the formation of the instability could be temperature dependent. 
Rewriting $f(r)$ from Eq. \eqref{metric} in terms of the temperature and horizon radius $r_0$, defined from the equation $f(r_0)=0$,  we obtain 
\bea
\label{metric2}
f(r) &=& L^2\left(\frac{4\pi T }{3 }\right)^2\frac{r^2}{r_0^2}\left(1-\frac{r_0^3}{r^3}\right) = L^2\alpha^2 \frac{r^2}{r_0^2}\left(1-\frac{r_0^3}{r^3}\right)
\eea
where  $r_0 = (ML^2)^{1/3}= \frac{4\pi L^2}{3} T$ is the radius of the event horizon and  $\alpha := 4\pi T/ 3$ is the temperature.
From this expression we see that as the temperature decreases, $|g_{tt}|$ decreases at fixed $r$ and therefore $|g^{tt}|$ increases, which means that the potential instability becomes stronger at low temperature. 

Following Ref. ~\cite{maeda1} we define an inverse radial coordinate $u=r_0/r$. Using this notation the metric takes the form
\bea
ds^2 = \frac{L^2\alpha^2}{u^2}\left(-h(u)dt^2 + d\Omega^2_{R^2}\right) +\frac{L^2du^2}{u^2h(u)} \text{~~with~~}
h(u) = 1-u^3\,.
\eea
In these coordinates the black hole horizon is located at $u=1$ and the boundary of AdS space is $u=0$.

In this paper we look at static field configurations in the bulk theory. 
One of the variables we use to study their stability is the free energy. 
We discuss the definition of the free energy below.
We start with the definition of the Euclidean action which is obtained from a Wick rotation of the Lorentzian action
\bea
\label{SE1}
S_E= -\int dx^4 \sqrt{-g} \, \left[ \mathcal{L}_{GR} + \mathcal{L}_{M} \right] = -S\,,
\eea
where we have separated the contributions due to gravity and matter by defining
\bea
\label{LGR}
&& \mathcal{L}_{GR} = \frac{c^4}{16\pi G} \left(R -2\Lambda \right) \\
\label{LM}
&& \mathcal{L}_{M} = -\frac{1}{4}F^{\mu\nu}F_{\mu\nu} -(D^{\mu}\psi)^{\dagger}D_{\mu}\psi  - m^2\psi^{\dagger}\psi\,.
 \eea
After the Wick rotation we have a new periodic Euclidean time coordinate with period, $\beta = 1/T$, inversely proportional to the temperature of the black hole \cite{Hertog}. For static, on-shell field configurations, the Euclidean action is proportional to the Gibbs free energy of the system\cite{BHThermo}. In the absence of time dependence, we can integrate over the time coordinate and write
\bea
\label{temper}
&& S_E =  \beta G = \frac{G}{T}
\eea
where the Gibbs free energy, $G$, involves an integral over the spatial coordinates only. In general, as stated in \cite{hartnoll2},  it is necessary to add extra terms to the action that depend on the boundary conditions imposed on the fields. In this paper we consider a simplified energy functional for the matter fields only which, for the boundary conditions we study in this paper, does not require any additional terms. We have verified this  explicitly by checking that our numerical results are independent of the grid spacing and size that is used to do the calculations (see Appendix \ref{A1}).
%


\subsection{Gradient flow}
\label{sec:GradientFlow}

Gradient flow is a  general analogue of the heat equation that describes the evolution (or flow) of fields with respect to a flow parameter, $\tau$  \cite{Headrick2006}.
We use a generalized index $I$ to represent a given field, and all internal indices associated with that field. The variable  $x$ is taken to represent all spacetime coordinates. For example, using this notation, the QCD gauge field $A^a_\mu(t,x^1\cdots x^d)$ would be denoted $\Phi^I(x)$, so that the Lorentz index $\mu$, and color index $a$, and the fact that we refer to a gauge potential (denoted with the letter $A$) are represented by one index $I$. A quark field $\psi(t,x^1\cdots x^d)$ would be represented similarly, but with a distinct index, for example as $\Phi^J(x)$.
The idea behind the gradient flow approach is to start with a set of fields $\Phi^I(x)$ and corresponding generating functional ${\cal F}[\Phi]$, and obtain a set of flow equations whose solutions move arbitrary initial field configurations along lines of steepest decent to an extremum of the generating functional.  

Given a functional ${\cal F}[\Phi]$ that depends on a set of fields $\Phi^I$, the flow equations are defined as
\bea
\label{flow1}
\frac{d\Phi^I(x)}{d\tau} = -G^{IJ}(\Phi) \frac{\delta {\cal F}[\Phi]}{\delta \Phi^J(x)}\,,
\label{eq:GeneralFlow}
\eea
where we have assumed the existence of a configuration space metric $G^{IJ}$, or equivalently, that the field space has a local, invertible and diffeomorphism invariant inner product that preserves any other symmetries of the system, 
\bea
\label{metric-def}
\langle  \delta\Phi  | \delta \Phi \rangle :=
   \int dx\, G_{IJ} \,\delta\Phi^I(x)\delta\Phi^J(x)\,.
\eea

We define an energy functional $\E$ in terms of the action by integrating out the time coordinate, similar to how the Gibbs free energy was defined in \eqref{temper}, so that the extrema of  $\E$ correspond to static solutions of the Euclidean action. We note that for on-shell configurations the energy functional is the Gibbs free energy $\E =G$. We use the energy functional $\E$ as the generating functional for our gradient flow equations.
Equation (\ref{flow1}) shows that, by construction, field configurations that extremize $\E$ will be fixed points of the flow and static equilibrium states of the superconductor. These fixed points can be stable, unstable or saddle points. Since the gradient flow finds the paths of steepest descent it is a natural tool for determining the stability of fixed points and studying the potential energy ``landscape.''

The flow parameter is arbitrary up to constant rescalings. Here we are interested in fixed points of the flow and the path that the system takes as it approaches equilibrium neither of which are affected by such rescalings.

The metric $G_{IJ}$ must respect the symmetries of the theory, and is normally read off from the gradient term in the free energy.  Specifically, consider a free energy of the form
\bea
\E = \int d^nx \left[ H_{IJ}(\Phi)D^\mu \Phi^I D_\mu \Phi^k+\hbox{lower order in derivatives}\right],
\eea
where $H_{IJ}$ depends only on the fields and not their derivatives. In this case it is natural to take $G_{IJ}=H_{IJ}$. 

For our matter fields 
\bea
\left(\Phi^I\right) &=&\threevector{A_{\mu}}{\psi}{\psi^\dagger},
\eea
we take $G_{IJ}$ to be
\bea
G_{IJ}=\left[\begin{tabular}{c c c}
$\sqrt{-g}g^{\mu\nu}$ & 0 & 0\\
$0$ & $0$& $\sqrt{-g}$\\
$0$ & $\sqrt{-g}$& $0$\\
\end{tabular}  
\right],
\eea
which is the simplest form for $G_{IJ}$ that still ensures the general covariance of the flow equations. In this paper we only consider the flow of the matter fields, so we do not need an inner product for the spacetime metric $g_{\mu\nu}$.

When the system contains different species of fields, as in our case,  one can consider the possibility that the different types of fields diffuse at different rates.  For example, one can recover the time-dependent Schrodinger equations considered in the standard Ginsburg-Landau model \cite{SuperFlow} by considering the following configuration space metric with the following block structure:
\bea
\left[
\begin{tabular}{c c c}
$k^{(1)}\sqrt{-g}g_{\mu\nu}$ & $0$ & 0 \\
$0$ & $0$& $k^{(2)}\sqrt{-g}$\\
$0$ & $k^{(2)}\sqrt{-g}$& $0$\\
\end{tabular}  
\right],
\eea
which reflects all the symmetries of the model. In the present case, where we consider bulk equations that come from an underlying supergravity theory, we assume for simplicity that the diffusion rates of the fields are the same, and consider the configuration space metric to come directly from the action.

From the differential flow equations (\ref{flow1}) we obtain field configurations for finite values of the flow parameter $\Phi^I(\tau,x)$ starting from specified initial conditions $\Phi^I(\tau = 0,x)$. We use the word solutions to refer to the single parameter family of field configurations $\Phi^I(\tau,x)$. 
The fixed points are defined as the configurations that satisfy $d\Phi^I(\tau,x)/d\tau = 0$. In our calculations, the fixed points that  minimize the energy are found numerically as the configurations reached at the end of the flow $\Phi^I(\tau \rightarrow \infty,x)$. The fixed points of the flow therefore satisfy
\bea
\frac{\delta \E[\Phi]}{\delta \Phi^J(x)} = 0,
\eea
which are the Euler-Lagrange equations of motion.

We note that the flow equations (\ref{flow1})  are covariant with respect to field reparametrizations of the form $\Phi^I \to \tilde{\Phi}^I(\Phi)$. 
However, even if the underlying theory is gauge and/or diffeomorphism invariant, the flow equations \eqref{flow1} are not necessarily gauge or diffeomorphism invariant.  Consequently, the gauge choice made for the initial field configuration is not in general preserved by the flow. This problem is addressed by adding a  deTurck term to the left-hand side of (\ref{flow1}) that  compensates the noninvariance of the right side.
Suppose the action (and therefore the energy) is invariant under the infinitesmal transformation,
\bea
\label{infin}
\delta{\Phi}^I(x) = K^I_{\alpha} (x) \xi^{\alpha}(x),
\eea
for a set of arbitrary parameters $\xi^\alpha(x)$
so that
\bea
\delta S = \int dx \frac{\delta S}{\delta \Phi^I(x)}K^I_{\alpha}(x)\xi^{\alpha}(x)=0\,.
\label{eq:GaugeInvariance}
\eea
The $K$'s generally involve differential operators and  functions of the fields. 
For example, in the case of Yang-Mills theory a gauge transformation of the vector potential is
\bea
\delta A^a_\mu(x) =  \partial_\mu \chi^a(x) - ig f^a_{bc} A^b_\mu(x) \chi^c(x)\,.
\eea
Using the notation in (\ref{infin}) we have
\bea
&& K^I_\alpha(x)  \rightarrow  \delta^a_c\partial_\mu  - ig f^a_{bc} A^b(x)_\mu \\
&& \xi^\alpha(x) \rightarrow \chi^c(x)
\eea
so that the index $I$ includes the indices $a$ and $\mu$, and the fact that the covariant derivative involves the gauge field $A$; the index $\alpha$ becomes $c$; and the  generic parameter $\xi$ is called $\chi$.
The general form of the flow equations, including the required deTurck term, is
\bea
\label{flow2}
\frac{d\Phi^I}{d\tau} +K^I_{\alpha} \xi^{\alpha} = - G^{IJ}\frac{\delta \E}{\delta \Phi^J} \,.
\label{eq:GeneralFlow2}
\eea
The role of the deTurck term is to ensure that any change in gauge along the flow can be compensated  by a  corresponding change in the parameters $\xi^{\alpha}$ .  

To see how the deTurck term preserves gauge invariance along the flow we write 
the rate of change of the energy with respect to the flow parameter $\tau$  as
\bea
\label{flow3}
\frac{d{\E}}{d\tau} = \int dx \,\frac{\delta{\E}}{\delta\Phi^J(x)}\frac{d\Phi^J(x)}{d\tau}
\eea
and substituting \eqref{flow2} into \eqref{flow3} we obtain
\bea
\label{flow4}
\frac{d{\E}}{d\tau} 
   &=&   \int dx \,\frac{\delta{\E}}{\delta\Phi^I(x)}\left( G^{IJ}\frac{d {\E}}{d\Phi^J(x)} -K^I_\alpha \xi^\alpha\right)\nonumber\\
    &=&  \int dx \,\frac{d{\E}}{d\Phi^I(x)}G^{IJ} \frac{d {\E}}{d\Phi^J(x)} \,.
\eea
where we have used \eqref{eq:GaugeInvariance} in the last step. 
The role of the deTurck term is illustrated explicitly in the context of the holographic superconductor  in the following section.
We also note that Eq. (\ref{flow4}) shows that if all the eigenvalues of the metric are positive, the energy is a monotonically decreasing function of the flow paramter $\tau$. In our case, since the eigenvalues of our metric are not all positive due to the term proportional to $g^{tt}$, the energy may not be monotonic. However, for the solutions we find by starting near the static vacuum fixed point, we find that the energy is monotonically decreasing. 

\section{Flow equations for the holographic superconductor}
\label{sec:FlowSuperconductor}
\subsection{Parametrization}
In this section we apply the formalism developed in the previous section to the action (\ref{action}). It is convenient to write the complex scalar field $\psi$ in terms of two real scalar fields, $p$ and $\omega$,
\bea
\psi = \frac{p}{q \sqrt{2}} e^{i\omega}.
\eea
We also rescale our vector field $\tilde{A}_{\mu} = q A_{\mu}$, which gives
\bea
\label{tilde-def}
\tilde{F}_{\mu\nu} = qF_{\mu\nu} = \partial_\mu \tilde{A}_\nu - \partial_\nu \tilde{A}_\mu  \text{~~and~~}  \tilde{D}_{\mu} = \partial_{\mu} - i \tilde{A}_{\mu}\,.
\eea
Using this notation the action in Eqs. \eqref{SE1}-\eqref{LM} becomes
\bea
S &=& \int dx^4 \sqrt{-g}\,( \mathcal{L}_{GR} + \frac{1}{q^2}\mathcal{L}_{M} ) \label{sepact}\\
\mathcal{L}_{GR} &=&\frac{c^4}{16\pi G} \left(R -2\Lambda \right) \\
\mathcal{L}_{M} &=& -\frac{1}{4}\tilde{F}^{\mu\nu}\tilde{F}_{\mu\nu} -\frac{1}{2}\left[\tilde{D}^{\mu}(p e^{i\omega})\right]^{\dagger} D_{\mu}(p e^{i \omega}) - m^2 p^2\, \nonumber \\
&=& \frac{-g^{\gamma\mu}g^{\delta\nu}}{2}(\partial_\mu \tilde{A}_\nu \partial\gamma \tilde{A}_\delta - \partial_\mu \tilde{A}_\nu \partial_{\delta} \tilde{A}_\gamma)  \nonumber \\
&&- \frac{g^{\mu\nu}}{2} p^2(\tilde{A}_\mu-\partial_{\mu}\omega)(\tilde{A}_\nu-\partial_{\nu}\omega) - \frac{g^{\mu\nu}}{2} \partial_\mu p \partial_\nu p  -\frac{1}{2}m^2 p^2 \,.\label{action-matter}
\eea

\subsection{Probe limit}
In this paper we consider the probe limit.
Mathematically we reach the probe limit by taking $q \rightarrow \infty$ in Eq. \eqref{sepact}, so that the gravity part of the action decouples from the matter part. 
Physically this means that we neglect backreaction of the gauge and scalar fields on the geometry itself. 
The basis of the approximation is the assumption that the terms containing the matter fields are negligible in the equations of motion for the metric components, which can then be solved to obtain the metric in (\ref{metric}). 
The equations of motion for the matter fields are then calculated after we substitute the metric in Eq. \eqref{metric} into the action, which means physically that we study the dynamics of the matter fields within a background metric.

In this limit we define an energy functional that includes only the matter part of the action
\bea
\label{Etilde}
\E &=& -\int dx^3 \sqrt{-g}\, \mathcal{L}_{M}\,,
\eea
which in the case of static fields only is
\bea
\E &=&\int d^3x \sqrt{-g} \left[\frac{g^{tt}}{2}(\partial^i \tilde{A}_t \partial_i \tilde{A}_t + \frac{1}{4}\tilde{F}^{ij}\tilde{F}_{ij} + \frac{g^{tt}}{2} \tilde{A}^2_t p^2 \right. \nonumber \\
&& \left. + \frac{p^2}{2} (\tilde{A}^i-\partial^{i}\omega)(\tilde{A}_i-\partial_i \omega) + \frac{1}{2} \partial^i p \partial_i p  +\frac{1}{2}m^2 p^2  \right],
\label{eq:MatterFreeEnergy}
\eea
where the $i,j$ indices are summed over spatial dimensions only. Since gradient flow moves along lines of steepest descent of $\E$, at any point along the flow, $\E$ is stationary in the directions orthogonal to it. In principle these directions can be integrated out to define an off-shell free energy and in a saddle point approximation this free energy is simply equal to the value of the action at that point along the flow \cite{Headrick2006}. We therefore refer to $\E$ as the energy of the system. We are primarily interested in the gradient flow between two fixed points. Since fixed points correspond to on-shell configurations, $\E$ is equal to the matter contribution to the Gibbs free energy at the start and end points of the flow.
From this point forward we drop the tildes that were introduced in Eqs. \eqref{tilde-def}.

Using \eqref{eq:MatterFreeEnergy} as the generating functional, the gradient flow equations for the matter fields are obtained from \eqref{flow2}. Before the inclusion of deTurck terms, the gradient flow equations for the matter fields are
\bea
\frac{\partial p}{\partial \tau} &=& \frac{1}{\sqrt{-g}}\partial_\nu(g^{\mu\nu}\sqrt{-g}\partial_\mu p) - p g^{\mu\nu}(A_\mu-\partial_\mu\omega)(A_\nu-\partial_\nu\omega) -m^2p \label{pflow} \\[2mm]
\frac{\partial \omega}{\partial\tau} &=& \frac{1}{\sqrt{-g}}\partial_\nu(g^{\mu\nu}\sqrt{-g}p^2(A_\mu - \partial_\mu\omega))
\label{omegaflow} \\[4mm]
\frac{\partial A_\mu}{\partial\tau} &=& \frac{g_{\mu\rho}}{\sqrt{-g}}\partial_\nu(\sqrt{-g}F^{\nu\rho}) - p^2 (A_\mu - \partial_\mu \omega) \,.\label{Aflow}
\eea
 The right sides of the flow equations are manifestly gauge invariant, but the left-hand side is not. 
 The action is invariant under the transformation $\{p,\omega,A_\mu\} \to \{p,\omega-\chi,A_\mu-\partial_\mu\chi\}$. 
Writing $\delta p=0$, $\delta \omega=-\chi$ and $\delta A_\mu = -\partial_\mu\chi$ Eq. (\ref{infin}) gives $K^p=0$ for the field $p$, and $K^\omega=-1$, $K^{A_\mu} =-\partial_\mu$ and $\xi = \chi$ for the remaining two fields. 
We see that the deTurck term $K^I_\alpha\xi^\alpha$ is 0 for the $p$ flow equation, which is what we expect since  $p$ is gauge invariant. The flow equations (\ref{flow2}) become
\bea
\frac{\partial p}{\partial \tau} &=& \frac{1}{\sqrt{-g}}\partial_\nu(g^{\mu\nu}\sqrt{-g}\partial_\mu p) - p g^{\mu\nu}(A_\mu-\partial_\mu\omega)(A_\nu-\partial_\nu\omega) -m^2p \label{pflowdt} \\[2mm]
\frac{\partial \omega}{\partial\tau}-\chi &=& \frac{1}{\sqrt{-g}}\partial_\nu(g^{\mu\nu}\sqrt{-g}p^2(A_\mu - \partial_\mu\omega))
\label{omegaflowdt} \\[4mm]
\frac{\partial A_\mu}{\partial\tau} - \partial_\mu \chi &=& \frac{g_{\mu\rho}}{\sqrt{-g}}\partial_\nu(\sqrt{-g}F^{\nu\rho}) - p^2 ( A_\mu - \partial_\mu \omega) \,.\label{Aflowdt}
\eea
At this stage $\chi$ is an arbitrary function.

A choice of the field $\omega$ can be interpreted as a gauge choice.
We can ensure that the flow in \eqref{omegaflowdt} preserves any initial value of $\omega$ by choosing
\bea
\label{deturck}
\chi &=& -\frac{1}{\sqrt{-g}}\partial_\nu(g^{\mu\nu}\sqrt{-g}p^2 (A_\mu - \partial_\mu \omega)) \,.
\eea
Making this choice our new set of flow equations is
\bea
\frac{\partial p}{\partial \tau} &=& \frac{1}{\sqrt{-g}}\partial_\nu(g^{\mu\nu}\sqrt{-g}\partial_\mu p) - p g^{\mu\nu}(A_\mu-\partial_\mu\omega)(A_\nu-\partial_\nu\omega) -m^2p \label{pflow2}\\
\frac{\partial \omega}{\partial\tau} &=& 0 \label{omegaflow2} \\
\frac{\partial A_\mu}{\partial\tau} &=& - \partial_\mu \left[\frac{1}{\sqrt{-g}}\partial_\beta(g^{\alpha\beta}\sqrt{-g}p^2 (A_\alpha - \partial_\alpha\omega))\right]+\frac{g_{\mu\rho}}{\sqrt{-g}}\partial_\nu(\sqrt{-g}F^{\nu\rho}) - p^2 ( A_\mu - \partial_\mu \omega)  \,.\nonumber\\\label{Aflow2}
\eea

\subsection{The boundary theory}
We are interested in how the gradient flow \eqref{pflow2} - \eqref{Aflow2} in the bulk changes the free energy and the condensate in the boundary superconducting theory. The AdS/CFT correspondence provides an equivalence between the Euclidean on-shell action in the bulk and the free energy of the boundary. This means that at the fixed points of the flow the free energy of the bulk, $\E$, is equivalent to the free energy of the superconductor. Away from the fixed points the AdS/CFT dictionary does not provide a relationship between the two free energies. The correspondence also links the scalar field in the bulk and an operator in the boundary theory that we take to represent the superconducting condensate. The dimension and expectation value of this operator is determined by the falloff of the scalar field $p$ at the AdS boundary $u=0$. 
The dimension of the operator depends on our choice of mass for the scalar field. Since the bulk metric is asymptotically AdS space the asymptotic behavior of the scalar field can be determined from the Klein-Gordon equation in ${\rm AdS}_{3+1}$. Generally there are two possible falloff rates $\Delta_{\pm}$,
\bea
\label{ass1}
p(u) &=& c_- u^{\Delta_-}\left(1+\mathcal{O}(u^2)\right)+ c_+ u^{\Delta_+}\left(1+\mathcal{O}(u^2)\right), \text{~~~where~~~} \\
\label{deldef}
\Delta_{\pm} &=& \frac{1}{2}\left(3\pm \sqrt{9+4m^2L^2}\right)\,,
\eea
but they are not always both normalizable. It is possible to consider tachyonic scalar fields with $m^2L^2<0$, and Breitenlohner and Freedman (BF) showed that ${\rm AdS}_{d+1}$ spacetime is stable if the scalar field  mass satisfies $m^2L^2>-d^2/4$ ~\cite{breit-freed}. Note that this bound is equivalent to the requirement that $\Delta_{\pm}$ is real. For masses near the BF bound, $-d^2/4 +1 > m^2L^2 > -d^2/4$ both terms in \eqref{ass1} are normalizable ~\cite{witten}, but if both coefficients are nonzero the theory is unstable in the asymptotic region ~\cite{herzog2007quantum}. We consider only the case $m^2L^2=-2$, where the asymptotic behavior of $p$ has a simple form
\bea
\label{aspec}
p(u) &=& c_1 u^{1} + c_2 u^2\,.
\eea
We can rewrite this equation in terms of the radial coordinate by recalling that we have defined $u =r_0/r$ with $r_0=\alpha L^2$. 
Defining $\langle{\cal O}_\pm\rangle = \alpha^{\Delta_\pm}c_\pm$, Eq. (\ref{aspec}) can be written
\bea
\label{ass2}
p\big(u(r)\big) = \frac{\langle{\cal O}_1\rangle}{r} + \frac{\langle{\cal O}_2\rangle}{r^2} + \cdots,
\eea
where $\langle{\cal O}_1\rangle$ and $\langle{\cal O}_2\rangle$ are the vacuum expectation values of operators in the boundary theory with dimension 1 and 2 respectively. In this case both terms are normalizable, but we confine our interest to $\langle{\cal O}_2\rangle$ by taking $c_1=0$.

We also introduce a finite charge density and chemical potential, which are obtained from the scalar potential $A_t$ in the boundary theory \cite{hartnoll2}. The motivation is that an additional scale is necessary to produce a superconducting instability at low temperatures.
For $u\rightarrow 0$ we write
\bea
\label{chem1}
A_t(u) = \mu - \bar{\rho} u + ...
\eea
where $\mu$ and $\bar{\rho}$ are, respectively, the chemical potential and charge density in the boundary theory (note that we use $\bar{\rho}$ for the charge density because $\rho$ is used as a radial coordinate when studying solutions with axial symmetry). 
The magnetic field in the boundary theory is obtained from 
\bea
\label{Bboundary}
\vec B = \vec\nabla \times \vec A(u=0,\vec x)\,
\eea
and the current can be obtained from the linear term in the expansion of the gauge potential around $u=0$ [see Eq. (\ref{Bpolar})].

\subsection{Gauge choice and {\it Ans\"atze}}

We consider three separate cases, with different symmetries on the boundary. \\
[10pt]
{\bf Spatially independent case}

We use coordinates in which $d\Omega^2_{R^2} = dx^2 +dy^2$. 
We simplify the equations by setting $A_u = A_x= A_y=\omega = 0$. We also take $A_t$ and $p$ to be functions of $u$ only. The complete set of conditions we impose is 
\bea
A_u&=&A_x=A_y=\omega=0 \label{cond1aa}\\
A_t&=&A_t(u)\,,~p=p(u) \label{cond2aa}\,.
\eea
The energy obtained from (\ref{eq:MatterFreeEnergy}) is
\bea
\E = \frac{1}{2} \int du \left[-\alpha (\partial_u A_t)^2 + \frac{\alpha^3 h(u)}{u^2} (\partial_u p)^2 + p^2\left(\frac{m^2 \alpha^3}{u^4} - \frac{\alpha A_t^2}{u^2 h(u)}\right)\right]\,.
\eea

The flow equations for the fields in \eqref{cond1aa} become $\dot{A}_u = \dot A_x = \dot A_y = \dot\omega=0$, where the dots denote derivatives with respect to the flow parameter. Thus the conditions in \eqref{cond1aa} are preserved by the flow. We also note that these conditions give $\chi = 0$,  which means that the deTurck term does not contribute.
From Eqs. \eqref{pflow2} and \eqref{Aflow2} we obtain the flow equations for the nonzero fields
\bea
\dot{A}_t &=& \frac{u^2 h}{L^2} \partial^2_u A_t - p^2 A_t \label{ratfaa} \\
\dot{p} &=& \frac{u^2}{L^2}\left[ u^2 \partial_u(\frac{h}{u^2}\partial_up) - p\left(\frac{-A^2_t}{\alpha^2 h} + \frac{L^2m^2}{u^2}\right)\right]\,.\label{rpfaa}
\eea
Since the metric depends only on the coordinate $u$, the conditions in \eqref{cond2aa} are also preserved by the flow. 
Equations \eqref{ratfaa} and \eqref{rpfaa} give a closed set of equations for two fields that depend on two spatial dimensions, $x$ and $u$, and the flow parameter.\\[10pt]
{\bf Translational symmetry}

We again consider coordinates in which $d\Omega^2_{R^2} = dx^2 +dy^2$ and  simplify the equations by setting $A_u = A_x= \omega = 0$. We now take $A_t$, $A_y$ and $p$ to be functions of $x$ and $u$ only, so our problem has a translational symmetry along the $y$ axis. The complete set of conditions we impose is 
\bea
A_u&=&A_x=\omega=0 \label{cond1}\\
A_t&=&A_t(u,x)\,,~A_y=A_y(u,x)\,,~p=p(u,x) \label{cond2}\,.
\eea
It is easy to see that with these conditions, the flow equations for the fields in \eqref{cond1} become $\dot{A}_u =\dot A_x = \dot\omega=0$,
which shows that the conditions in \eqref{cond1} are preserved by the flow. We also note that these conditions again give $\chi = 0$, so that the deTurck term
does not contribute.
The energy (\ref{eq:MatterFreeEnergy}) in this case is
\bea
\E = \frac{1}{2} \int du dx  \left[-\alpha (\partial_u A_t)^2 + \alpha h (\partial_u A_y)^2 - \frac{(\partial_x A_t)^2}{h \alpha} + \frac{(\partial_x A_{y})^2}{\alpha} + \frac{\alpha}{u^2} (\partial_x p)^2  \right. \nonumber \\ \left. + \frac{\alpha^3 h(u)}{u^2} (\partial_u p)^2 + p^2\left(\frac{m^2 \alpha^3}{u^4} + \frac{\alpha A_{y}^2}{u^2}- \frac{\alpha A_t^2}{u^2 h(u)}\right)\right]\,.
\eea

From equations \eqref{pflow2} and \eqref{Aflow2} we obtain the flow equations for the nonzero fields
\bea
\dot{A}_t &=& \frac{u^2 h}{L^2} \partial^2_u A_t + \frac{u^2 h}{L^2 \alpha^2} \partial^2_x A_t - p^2 A_t, \label{ratf} \\
\dot{A}_y &=& \frac{u^2}{L^2}\partial_u(h\partial_u(A_y)) + \frac{u^2}{L^2\alpha^2}\partial^2_x A_y - p^2  (A_y),\label{rayf} \\
\dot{p} &=& \frac{u^2}{L^2}\left[\frac{1}{\alpha^2} \partial^2_x p + u^2 \partial_u(\frac{h}{u^2}\partial_up) - p\left(\frac{-A^2_t}{\alpha^2 h}+\frac{(A_y)^2}{\alpha^2} + \frac{L^2m^2}{u^2}\right)\right]\,.\label{rpf}
\eea
Since the metric depends only on the coordinate $u$, the right sides of these equations are independent of $y$ and $t$, which shows
that the conditions in \eqref{cond2} are also preserved by the flow. 
Equations \eqref{ratf} - \eqref{rpf} give a closed set of equations for three fields that depend on two spatial dimensions, $x$ and $u$, and the flow parameter.

We note that using these coordinates the magnetic field in the boundary theory is obtained from Eq. (\ref{Bboundary}) as
\bea
\label{Bcart}
B(x) = \partial_x A_y(u=0,x) \,.
\eea

{\bf Axial symmetry} 

We also consider coordinates where $d\Omega^2_{R^2} = d\rho^2 +\rho^2 d\theta^2$. 
One motivation is that rotational symmetry allows us to study completely localized solutions that could be created in a lab. When working with axial symmetry one typically looks for solutions where the phase of the complex scalar field can be written $\omega = n \theta$, and $n$ is interpreted as an integer winding number. The value of the winding number is an important property of vortex solutions and leads to flux quantization in superconductors. We once again take $\omega = 0$, which can be thought of as before as a gauge choice, but since we impose axial symmetry on the remaining fields, it also restricts us to solutions with zero winding number. 
We further take $A_u = A_\rho = 0$, and assume that our remaining fields are functions of $u$ and $\rho$ only. The complete set of conditions we use is 
\bea
A_u&=&A_\rho = 0, \label{cond1b}\\
\omega &=& n \theta \,,~n=0,\label{winding}\\
A_t&=&A_t(u,\rho)\,,~A_\theta=A_\theta(u,\rho)\,,~p=p(u,\rho) \label{cond2b}\,.
\eea
The energy (\ref{eq:MatterFreeEnergy}) becomes
\bea
\E = \frac{1}{2} \int du \,dr\, r\,  \left[-\alpha (\partial_u A_t)^2 + \alpha h (\partial_u A_\theta)^2 - \frac{(\partial_r A_t)^2}{h \alpha} + \frac{(\partial_r A_{\theta})^2}{\alpha} + \frac{\alpha}{u^2} (\partial_r p)^2  \right. \nonumber \\ \left. + \frac{\alpha^3 h(u)}{u^2} (\partial_u p)^2 + p^2\left(\frac{m^2 \alpha^3}{u^4} + \frac{\alpha A_{\theta}^2}{r u^2}- \frac{\alpha A_t^2}{u^2 h(u)}\right)\right]\,.
\eea

We note that as before the flow equations give $\dot{A}_u =\dot{A}_\rho = \dot{\omega}=0$ so that the conditions \eqref{cond1b} and \eqref{winding} are preserved by the flow, and we also have again from (\ref{deturck}) that $\chi =0$, which means that the deTurck term does not contribute. The remaining flow equations are
\bea
\dot{A}_t &=& \frac{u^2 h}{L^2} \partial^2_u A_t + \frac{u^2}{L^2 \alpha^2}\left(\partial^2_{\rho} A_t +\frac{1}{\rho}\partial_{\rho}A_t \right) - p^2 A_t, \label{ratfb} \\
\dot{A}_{\theta} &=& \frac{u^2}{L^2}\partial_u(h\partial_u(A_{\theta})) + \frac{u^2}{L^2\alpha^2}\left[\partial^2_{\rho} A_{\theta}-\frac{1}{\rho}\partial_{\rho} A_{\theta}\right]  - p^2  A_{\theta},\label{rayfb} \\
\dot{p} &=& \frac{u^2}{L^2}\left[\frac{1}{\alpha^2} \left(\partial^2_{\rho} p +\frac{1}{\rho}\partial_{\rho} p \right) + u^2 \partial_u \left(\frac{h}{u^2}\partial_u p \right) - p\left(\frac{-A^2_t}{\alpha^2 h}+\frac{A_{\theta}^2}{\alpha^2 \rho^2} + \frac{L^2m^2}{u^2}\right)\right].\label{rpfb}
\eea

The azimuthal component of the vector potential of the boundary theory can be written [see Eq. (\ref{Bboundary})] as
\bea
\label{Bpolar}
A_{\theta}(u) = \frac{B}{2}\rho^2 + J_{\theta} u + ...
\eea
where $B$ is the magnetic field and $J_\theta$ is an azimuthal current.

\section{Solutions to the Gradient Flow} 
\label{GF-solutions}

To solve our gradient flow equations we need to proceed numerically, starting from a specified  initial configuration for the fields. Since we are particularly interested in studying the flow between two fixed points, we consider the vacuum (hairless black hole) configuration which has a simple analytic form
\bea
A_t = \mu(1-u) \,,~~
A_y = Bx \,,~~
p = 0 \label{pert-trans}
\eea
for the translationally symmetric case, and 
\bea
A_t = \mu(1-u) \,,~~
 A_\theta = \frac{B}{2} \rho^2\,,~~
p = 0 \label{pert-axial}
\eea
for the axially symmetric case.
It is easy to verify that these configurations are fixed points of the flow. 
From Eqs. \eqref{Bcart} and \eqref{Bpolar} we see that $B$ is an arbitrary constant external magnetic field.
We start the flow from the vacuum configurations in (\ref{pert-trans}) or (\ref{pert-axial}) with the addition of a small perturbation of the scalar field, $\delta p$.
We employ a simple explicit finite difference method with a forward difference in flow time and centered difference in spatial coordinates. Further details about the numerical method and boundary conditions can be found in Appendix \ref{A1}.

One feature of the gradient flow method is that if our perturbation satisfies $\delta p(u=0) = 0$, then $\dot{A}_{\mu}(u=0) = 0$.   This is true for all of our {\it Ans\"{a}tze} and can be seen directly from Eqs \eqref{ratfaa}, \eqref{ratf}, \eqref{rayf}, \eqref{ratfb}, and \eqref{rayfb}. We have therefore that in the boundary theory, the chemical potential $\mu = A_t(0)$ and magnetic field $B = \partial_x A_y(0,x)$ or $B=\frac{1}{\rho}\partial_\rho A_\theta(0,\rho)$, are specified by our initial configuration and constant along the flow.

A more general statement is that using our method the gauge field on the boundary $A_{\mu}(0,\vec x)$, and all its derivatives with respect to boundary coordinates, are fixed by our initial configuration and unchanged along the flow. 
This means that the boundary theory does not have a dynamical gauge field, which corresponds to a limit where the superconductor is equivalent to a superfluid. It is possible to make the gauge field dynamical by including an additional boundary term in the bulk action and considering a different type of boundary condition on $A_{\mu}$ \cite{Salvio}. While a dynamical gauge field is important for many superconductor phenomena such as the Meissner effect, a fixed background is sufficient to study how gradient flow in the bulk creates a corresponding flow in the boundary, and the extension of the flow to more complicated systems is straightforward.
We comment that in the boundary theory, while derivatives with respect to boundary coordinates $\partial_x A_{\mu}(0,\vec x)$ and $\partial_y A_{\mu}(0,\vec x)$ are fixed (as explained above), 
derivatives with respect to the AdS coordinate $\partial_u A_{\mu}(u,\vec x)\big|_{u=0}$ are not fixed [see Eqs. \eqref{chem1}, \eqref{Bpolar}].

We are interested in how the flow alters the condensate operator and the energy in the boundary theory. With respect to the energy, the quantity of physical interest is the change in the energy along the flow, relative to the energy of the vacuum state $\Delta E = {\cal E}-{\cal E}_{\rm vac}$. Using this normalization we find that when the vacuum solution is unstable the quantity $\Delta E$ moves from an initial value of 0 into negative values.
This type of behavior is typical in systems that exhibit spontaneous symmetry breaking, where a false vacuum decays into a more stable (but less symmetric) configuration.

In numerical calculations we set $L = 1$, which is equivalent to using the AdS radius as a length scale in all dimensional quantities (including the flow parameter). We also set the chemical potential $\mu = 1$. This is equivalent to defining new $\rho$ (or $x,y$) coordinates and absorbing $\mu$ into the parameter $\alpha$ in the flow equations as follows:
\bea
&& A_t \rightarrow  \tilde{A}_t =\frac{A_t}{\mu},\\
&& \rho \rightarrow \tilde{\rho} = \mu\rho, \\
&& \alpha \rightarrow \tilde{\alpha} = \frac{\alpha}{\mu} = \frac{4\pi T}{3\mu},\\
&& B \rightarrow \tilde{B} = \frac{B}{\mu^2}\,.
\eea
In the following we omit the tilde’s but understand that setting $\mu = 1$ in the solution implies we are in fact referring to the above rescaled quantities in our results.
We vary the parameter $\alpha=4\pi T/3$ (or equivalently the temperature) and the magnetic field $B$.

\subsection{Spatially independent solutions}
\label{spat-indep}

\begin{figure}[ht]
\begin{subfigure}[h]{0.49\textwidth}
\includegraphics[width=\textwidth]{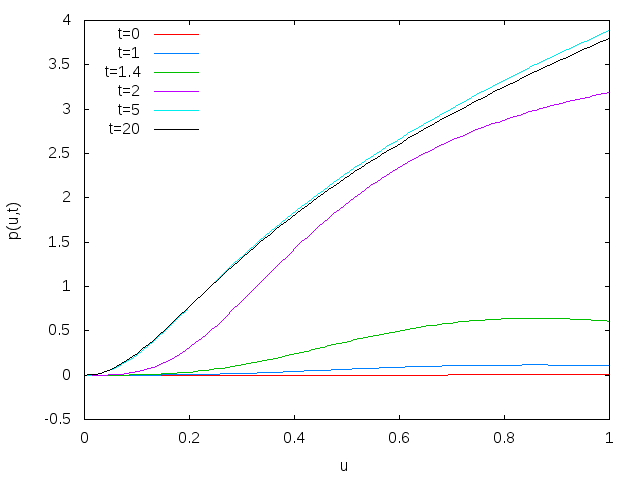}
\caption{For $\alpha^2 = 0.01$ a perturbation away from\\ $p=0$ is unstable.\label{unstable}}
\end{subfigure}
\begin{subfigure}[h]{0.49\textwidth}
\includegraphics[width=\textwidth]{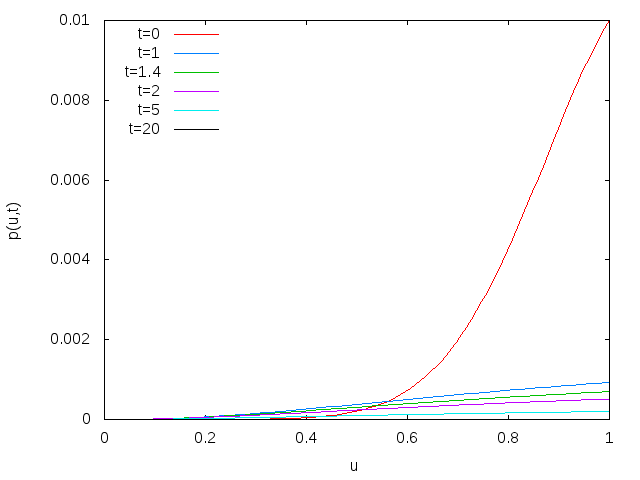}
\caption{For $\alpha^2 = 0.1$ a perturbation away from $p=0$ returns to zero.\label{stable}}
\end{subfigure}
\caption{The evolution of the scalar field $p(u)$, at several values of $\tau$ along the flow for spatially independent fields.\label{stab}}
\end{figure}

By varying  $\alpha = 4\pi T/3$ we can determine the critical value below which the vacuum becomes unstable. We start with an initial perturbation of the scalar field given by
\bea
\delta p(u) = 10^{-3}\times u^2 e^{-(u-1)^2}\,.
\eea
Figure \ref{stab} shows how this initial perturbation evolves for two different values of the temperature.
In Fig. 1(\subref{unstable}) the bulk scalar field moves away from the (false) vacuum, and in Fig. 1(\subref{stable}) we see that at higher temperature the perturbed system returns to the vacuum configuration. 

\begin{figure}[ht]
\begin{subfigure}[h]{0.49\textwidth}
\includegraphics[width=0.9\textwidth]{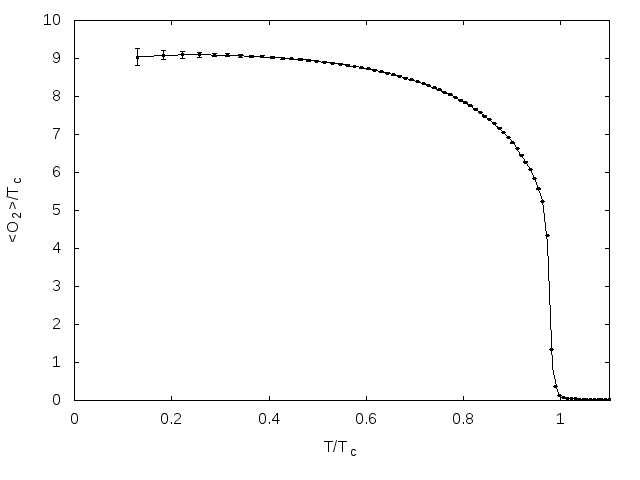}
\caption{$\langle O_2 \rangle$ as a function of $T$ \label{HFig1}}
\end{subfigure}
\begin{subfigure}[h]{0.49\textwidth}
\includegraphics[width=0.9\textwidth]{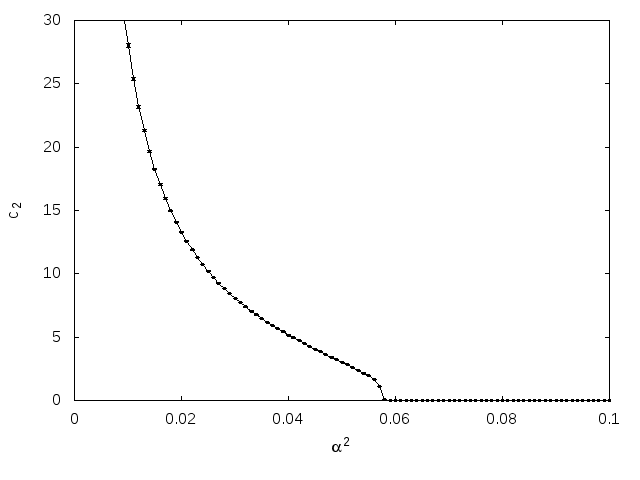}
\caption{$c_2$ as a function of $\alpha^2$ \label{Crit}}
\end{subfigure}
\caption{The dependence of the condensate on temperature for spatially independent fields.}
\end{figure}

To compare with the results of ~\cite{hartnoll1}, we recall $\langle{\cal O}_\pm\rangle = \alpha^{\Delta_\pm}c_\pm$ and plot $\langle O_2\rangle$ versus $T$. 
The parameter $c_2$ is obtained from a second derivative of the scalar field [see Eq. (\ref{ass1})], which is calculated using a finite difference formula on the first three data points. The error bars are obtained by comparing with the value extracted from the first, third and fifth data points.
The result is shown in Fig. 2(\subref{HFig1}), and agrees well with Fig. 1(b) in Ref. ~\cite{hartnoll1}.  
In Fig. 2(\subref{Crit}) we show the same data using different variables: using $\alpha = 4\pi T/3 = r_0/L^2$ we plot $c_2$ versus $\alpha^2$. From  Fig. 2(\subref{Crit}) we see that the critical temperature is approximately $\alpha^2_c \approx 0.059$.\\

\begin{figure}[ht]
\begin{subfigure}[h]{0.49\textwidth}
\includegraphics[width=0.9\textwidth]{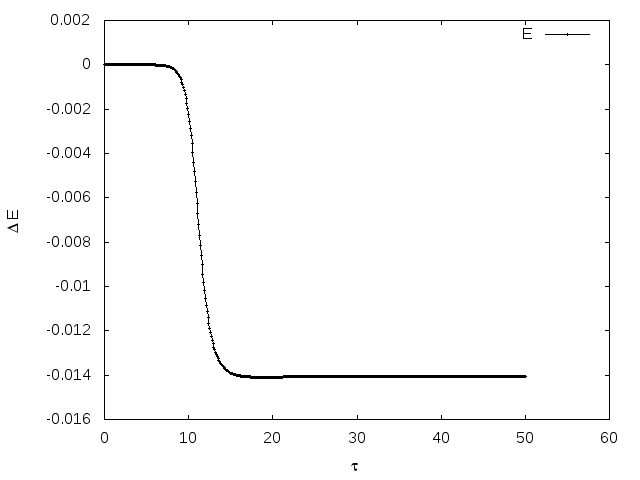}
\caption{$\Delta E$ as a function of $\tau$ for spatially independent fields \label{Etau}}
\end{subfigure}
\begin{subfigure}[h]{0.49\textwidth}
\includegraphics[width=0.9\textwidth]{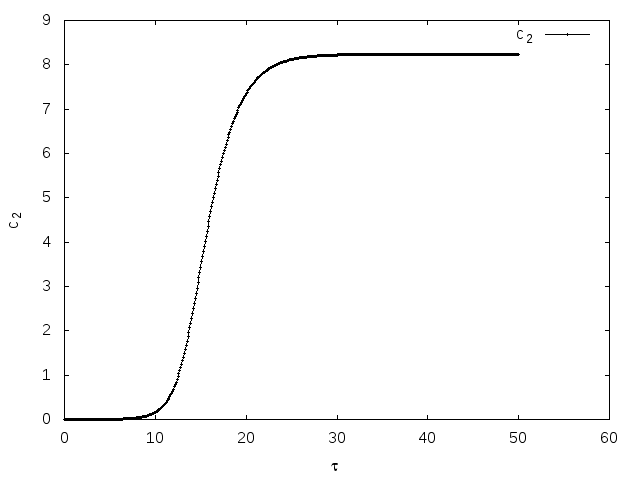}
\caption{$c_2$ as a function of $\tau$\label{c2tau}}
\end{subfigure}
\begin{subfigure}[h]{0.49\textwidth}
\includegraphics[width=0.9\textwidth]{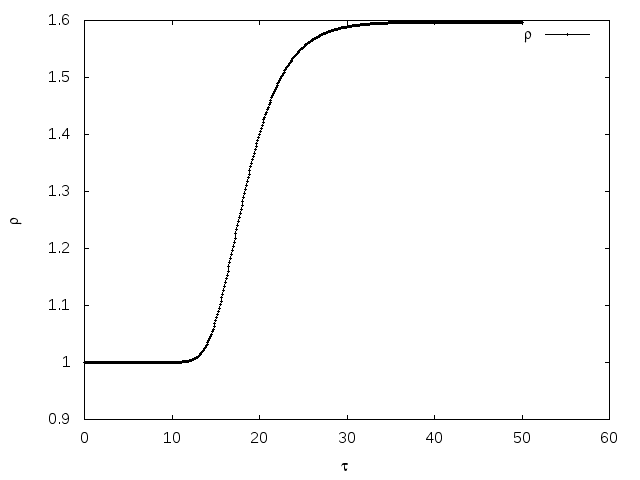}
\caption{$\bar{\rho}$ as a function of $\tau$ \label{cdtau}}
\end{subfigure}
\caption{The flow of quantities in the boundary theory for $\alpha^2=0.03$.\label{vstau}}
\end{figure}

We also want to study the evolution of physical quantities in the boundary theory as a function of the flow parameter. 
In Fig. \ref{vstau} we show the evolution of the energy, charge density, and condensate operator with the flow parameter.

\begin{figure}[ht]
\includegraphics[width=0.49\textwidth]{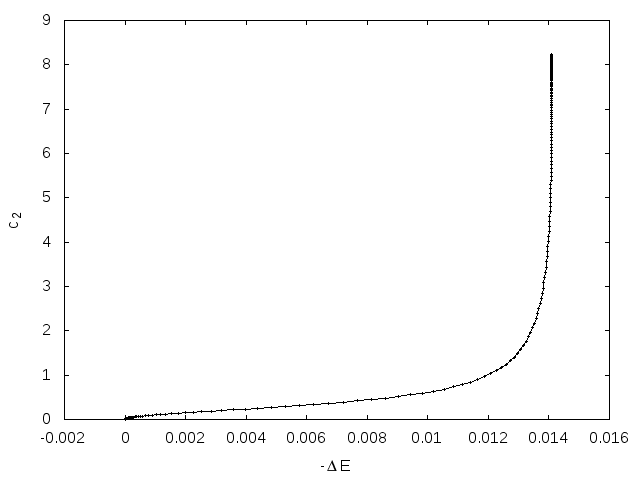}
\caption{$c_2$ as a function of $-\Delta E$ for spatially independent fields at $\alpha^2=0.03$. \label{Ec2}}
\end{figure}

Since the energy is a decreasing quantity along the flow between the vacuum fixed point and the scalar hair fixed point, we can look at the quantities in the boundary theory as functions of the bulk energy instead of the flow parameter, which has no straightforward physical interpretation. 
Figure \ref{Ec2}\,  shows how the condensate changes as the energy of the system decreases. 

\subsection{Translationally symmetric dark solitons}

We notice that the equations used for the spatially independent solutions $p_{\rm si}$ are invariant under the transformation $p \rightarrow -p$. This suggests that there may be stable fixed points where $p$ is an odd function of $x$ such that 
$\lim_{x\to\pm\infty} p(x) = \pm p_{\rm si}$.
%
Such configurations are called dark solitons \cite{keranen}. We can consider only the portion of the soliton where $p>0$ by looking at only $x>0$ and enforcing the condition that $p(u,0)=0$. 
We need to start from a perturbation that satisfies this condition and therefore we choose
\bea
\delta p(u,x) = 10^{-2}\times u^2 e^{-10(u-1)^2}\tanh{(5x)}.
\eea

This flow can be interpreted as either a full soliton for $-\infty < x < \infty$ where $p$ is antisymmetric around $x=0$, or as a solution for $x>0$ where $x=0$ is an interface with a fixed vacuum solution for $x<0$. In the boundary theory we can interpret the second case as an interface between a superconductor (for $x>0$) and a normal material (for $x<0$).
 In the Ginzburg-Landau (GL) theory of superconductivity there is an exact solution for the order parameter $\phi_{Gl}$
\bea
\label{Gl}
\phi_{Gl} = \phi_{\infty} \tanh\left(\frac{x}{\sqrt{2}\xi}\right),
\eea
where $\phi_{\infty}$ is the value of the order parameter in the pure superconducting phase, and $\xi$ is the coherence length. 

\begin{figure}[ht]
\begin{subfigure}[h]{0.49\textwidth}
\includegraphics[width=0.9\textwidth]{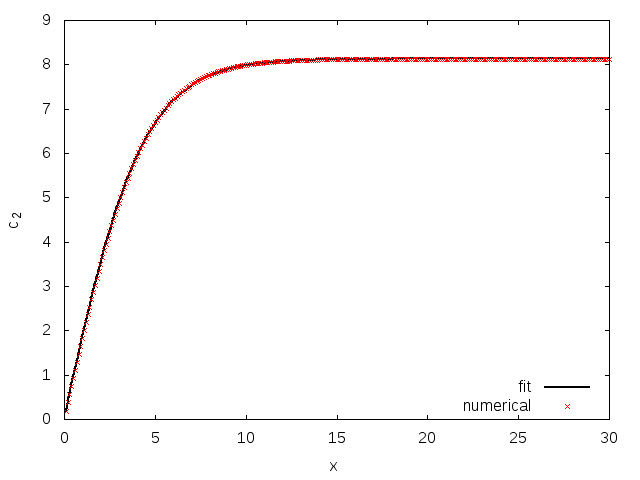}
\caption{The fit of $c_2(x)$ to a tanh function for $\alpha^2=0.03$.
\label{tanhfit}}
\end{subfigure}
\begin{subfigure}[h]{0.49\textwidth}
\includegraphics[width=0.9\textwidth]{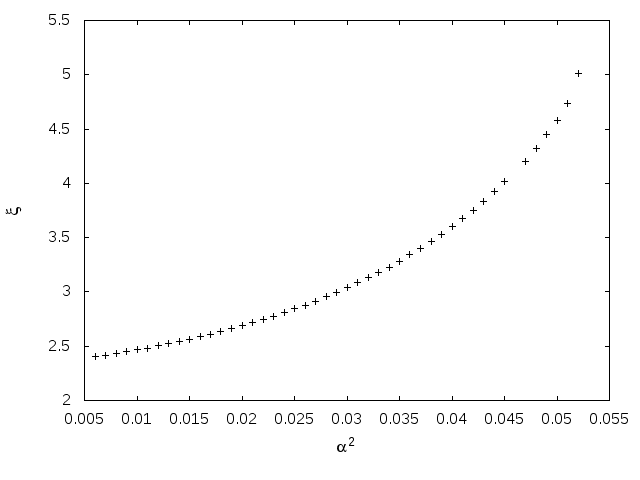}
\caption{The coherence length as a function of $\alpha^2$ \label{xivst}}
\end{subfigure}
\caption{The dependence of the coherence length on temperature for the dark soliton fixed points}
\end{figure}

Although we would like a solution for all $x>0$, to perform the numerical calculation we need to introduce a cutoff. We would like our cutoff to be large enough that our fields are constant for $x>x_{\rm max}$. From Eq. \eqref{Gl} we find that for $x_{\rm max}\approx 5\xi$ we have $p(x=x_{\rm max})>0.998 \times p(x=\infty)$. We use a cutoff  $x_{\rm max}=30$ and we find $x_{\rm max}/\xi > 5.9$ for $T\approx 0.93T_c$.
Our numerical results indicate that the boundary operator has a similar $x$ dependence, and we can therefore fit to a function of the same form as Eq. (\ref{Gl}) to determine the coherence length of the holographic superconductor. The fit for $\alpha^2=0.03$ is shown in Fig. 5(\subref{tanhfit}). We can see in Fig. 5(\subref{xivst})\, that, as expected, the coherence length diverges proportional to $1/\sqrt{1-\alpha/\alpha_c} = 1/\sqrt{1-T/T_c}$ as we approach the critical temperature.
We note that due to this divergence, the condition $x_{\rm max}/\xi >1$ cannot be satisfied very close to the critical temperature and for this reason we have considered only $T\leq 0.93 T_c$.

\begin{figure}[ht]
\includegraphics[width=0.49\textwidth]{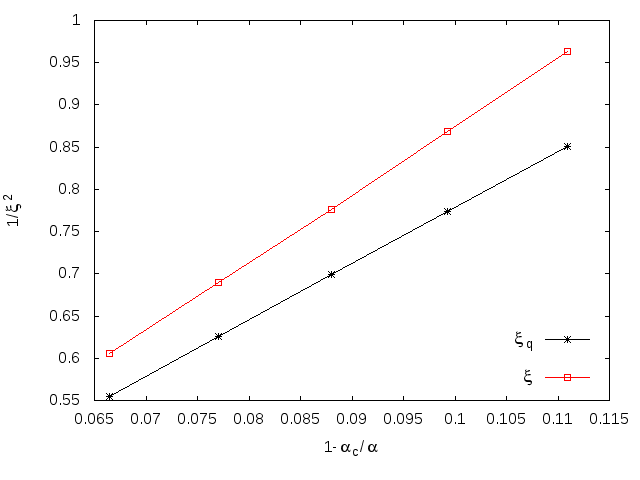}
\caption{The two coherence lengths $\xi$ (for the scalar field) and $\xi_{q}$ (for the charge density) as a function of the deviation from the critical temperature for soliton solutions. \label{cohlengths}}
\end{figure}

In the GL theory the charge density is proportional to $\phi^2$,
\bea
\bar{\rho} \propto \phi^2 \propto \tanh^2\left(\frac{x}{\sqrt{2}\xi}\right) = 1 - {\rm sech}^2\left(\frac{x}{\sqrt{2}\xi}\right).
\eea
 We can therefore find a characteristic length for the charge density $\bar{\rho}(x)$ by fitting  to ${\rm sech}^2(\frac{x}{\sqrt{2}\xi_{q}}).$ Contrary to what  we expect from  GL theory, we find that the two length scales are different. The difference between $\xi$ and $\xi_{q}$ increases as we move further from the critical temperature, as can be seen in Fig. \ref{cohlengths}. This result agrees with what was found in \cite{keranen} (apart from a difference in how coherence length is defined in the GL compared to Gross-Pitaevskii equations).

\begin{figure}[ht]
\includegraphics[width=0.49\textwidth]{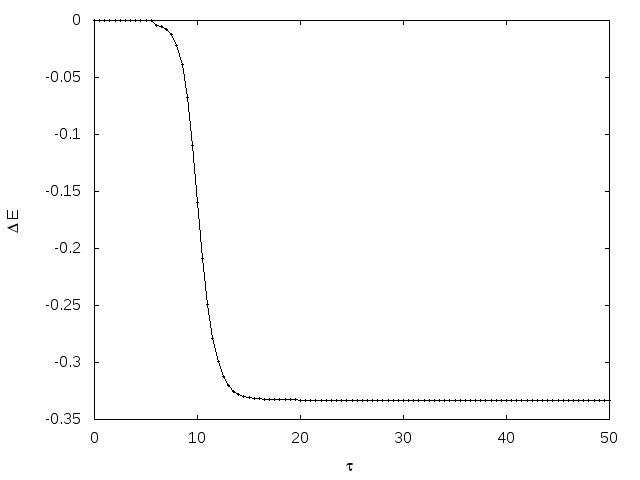}
\caption{The energy ($\Delta E$) as a function of the flow parameter ($\tau$) with $\alpha^2 = 0.03$ for the soliton solution. \label{ESoli}}
\end{figure}
\begin{figure}[ht]
\includegraphics[width=0.49\textwidth]{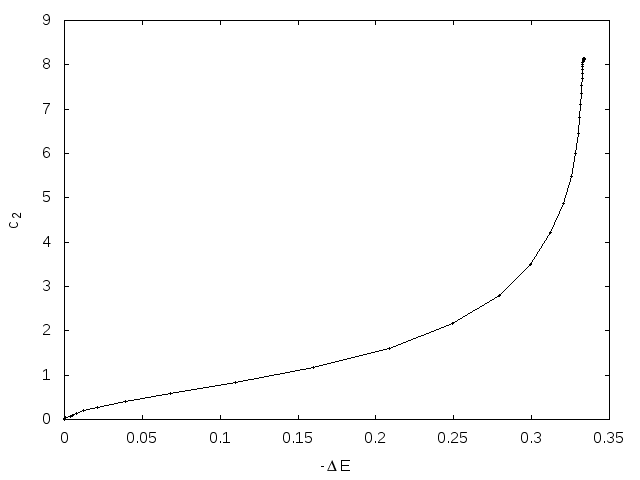}
\caption{The operator ($c_2(x=30)$) as a function of the energy ($\Delta E$) with $\alpha^2 = 0.03$ for the soliton solution. \label{Ec2Soli}}
\end{figure}
We can calculate the energy density and the boundary operator for the soliton configuration as functions of $x$ and $\tau$. 
In Fig. \ref{ESoli} we show the energy density integrated over $x$ as a function of the flow parameter.
Since $\Delta E$ decreases monotonically with $\tau$, we can look at the evolution of the boundary operator as a function of $|\Delta E|$ instead of $\tau$.
We have shown that our $x$ cutoff is larger than typical values of the coherence lengths [see Fig. 5(\subref{xivst})], which means that field configurations are approximately constant at large $x$ [see for example Fig. 5(\subref{tanhfit})].
We therefore expect $c_2(x=x_{\rm max}) \approx c_2(x \to \infty)$, and that the soliton solution at large $x$ should be close to the spatially independent solution we considered in Sec. \ref{spat-indep}. In Fig. \ref{Ec2Soli} we plot the evolution of the boundary operator for large $x$, $c_2(x=x_{max})$ , as a function of the energy.
In the soliton case we have $\Delta E_{\rm Sol} \approx -0.334$, and if we calculate the energy of the spatially independent case on the same interval ($0\leq x \leq 30$) we find $\Delta E_{\rm Ind} \approx -0.395.$ 
These energies depend on the cutoff ($x_{\rm max}$), but the difference $\Delta E_{\rm Sol}-\Delta E_{\rm Ind}$ will be independent of the cutoff and can be thought of as a measure of the effect of the soliton. 
 \\

\subsection{Droplet solutions}
The final case we consider is configurations with nonzero magnetic field. 
The interesting feature of these solutions is that the magnetic field can limit the formation of the scalar hair/condensate. 
There is a critical magnetic field $B_c$ above which the boundary operator does not condense. 
For $B<B_c$ the condensate forms only in a localized region on the boundary, and these localized solutions are called droplet solutions ~\cite{maeda2}. 
For the droplet solutions, the initial perturbation does not need to depend on $x$, so we use the perturbation 
\bea
\delta p(u) = 10^{-2}\times u^2 e^{-10(u-1)^2}.
\eea
\begin{figure}[ht]
\includegraphics[width=0.49\textwidth]{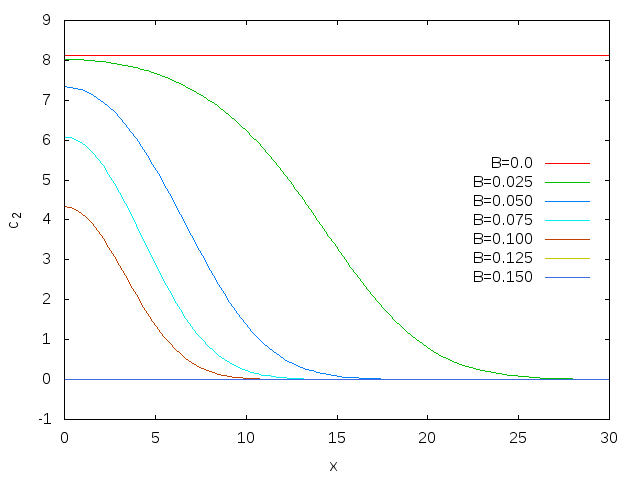}
\caption{$c_2(x)$ for several values of the magnetic field, and $\alpha^2 = 0.03$, for a translationally symmetric droplet. As $B$ increases $c_2(x)$ becomes smaller and more localized. \label{Bc1}}
\end{figure}

\subsubsection{Translationally symmetric droplets}
\begin{figure}[ht]
\includegraphics[width=0.49\textwidth]{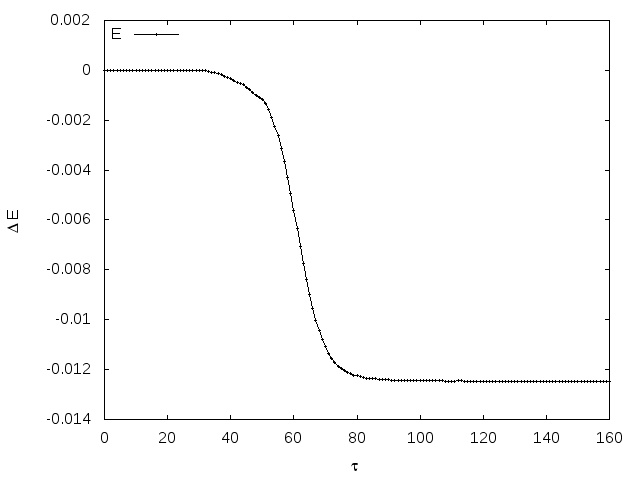}
\caption{The energy ($\Delta E$) as a function of the flow parameter ($\tau$) with $\alpha^2 = 0.03$ and $B=0.075$ for a translationally symmetric droplet.\label{EDrop}}
\end{figure}

\begin{figure}[ht]
\includegraphics[width=0.49\textwidth]{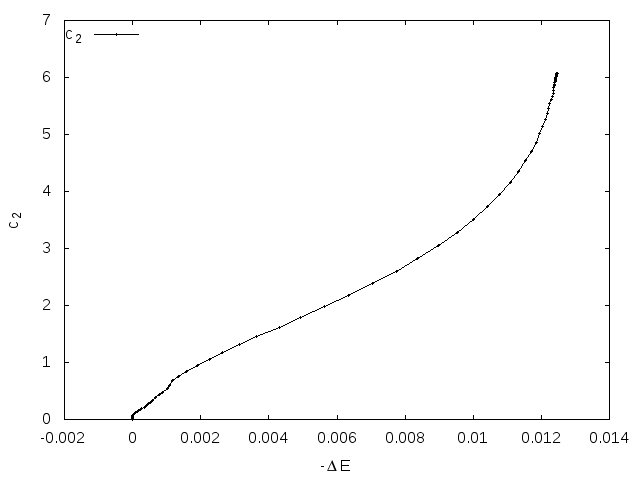}
\caption{The energy ($\Delta E$) as a function of the operator at the center of the droplet ($c_2(x=0)$) with $\alpha^2 = 0.03$ and $B=0.075$ for a translationally symmetric droplet.\label{Ec2Drop}}
\end{figure}

In Fig. \ref{Bc1} we show how the magnetic field alters the boundary operator $c_2(x)$. When $B\to 0$ we recover the spatially independent solution [see Fig. 3(\subref{c2tau})].
As $B$ approaches the critical value from below, the droplets become narrower and shorter. 
In Fig. \ref{EDrop} we see how the energy evolves as the flow moves the fields from a small perturbation of the vacuum solution towards a localized droplet of scalar hair. In Fig. \ref{Ec2Drop} we  plot the energy versus the value of the operator $c_2$ at the center of the droplet. We notice in Fig. \ref{EDrop} that when $\Delta E$ is small there are noticeable fluctuations due to numerical error, which influence the low $\Delta E$ behavior in Fig. \ref{Ec2Drop}.

\subsubsection{Axially symmetric droplet solutions}

\begin{figure}[ht]
\includegraphics[width=0.49\textwidth]{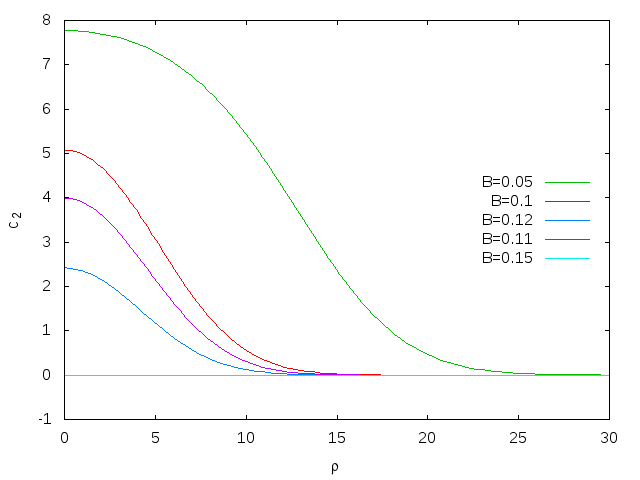}
\caption{The operator $c_2(x)$ for several values of the magnetic field, and $\alpha^2 = 0.03$, for an axially symmetry droplet. \label{Bcrot}}
\end{figure}

The droplet solutions do not require translational symmetry. In this section we start from the same initial perturbation but instead enforce axial symmetry along the flow. This leads to droplet solutions that are entirely localized (as condensate in the boundary theory and as scalar hair on the black hole horizon). 
Our results are similar to those found in \cite{albash}, with the important difference that we have a constant magnetic field on the boundary. As discussed at the beginning of Sec. \ref{GF-solutions}, this is a characteristic of the gradient flow method. 
We study the flow up to a maximum radius $\rho=30$, for several different values of the magnetic field. The final configurations for $c_2$ are shown in Fig. \ref{Bcrot}. 

\begin{figure}[ht]
\includegraphics[width=0.49\textwidth]{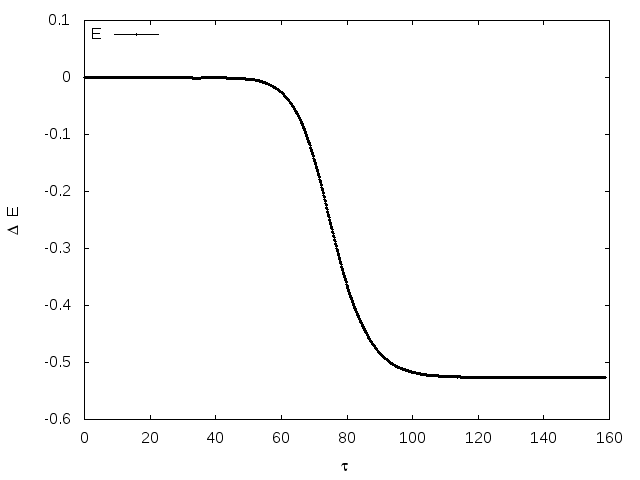}
\caption{The energy ($\Delta E$) as a function of the flow parameter ($\tau$) with $\alpha^2 = 0.03$ and $B=0.1$ for an axially symmetric droplet.\label{EAx}}
\end{figure}

\begin{figure}[ht]
\includegraphics[width=0.49\textwidth]{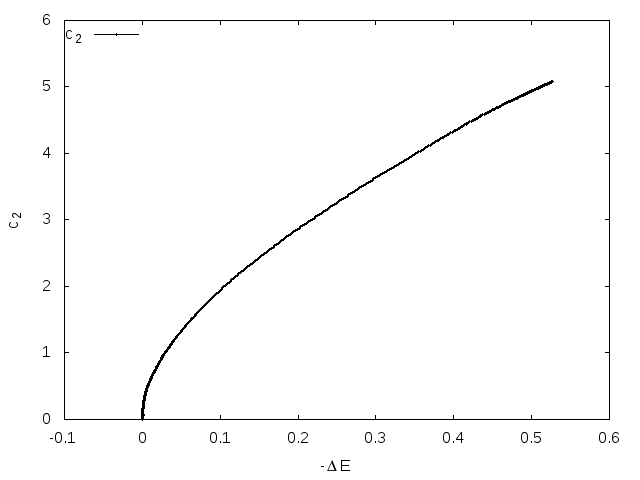}
\caption{The operator  at the center of the droplet ($c_2(\rho=0)$) as a function of the energy ($\Delta E$) with $\alpha^2 = 0.03$ and $B=0.1$ for an axially symmetric droplet. \label{Ec2Ax}}
\end{figure}

In Figs. \ref{EAx} and \ref{Ec2Ax} we show the energy for the case where $\alpha^2 = 0.03$ and $B=0.1$. In the first figure we show how the energy evolves as a function of flow time, and in the second we plot the energy versus the value of the boundary operator at $\rho=0$.


\begin{figure}[ht]
\includegraphics[width=0.49\textwidth]{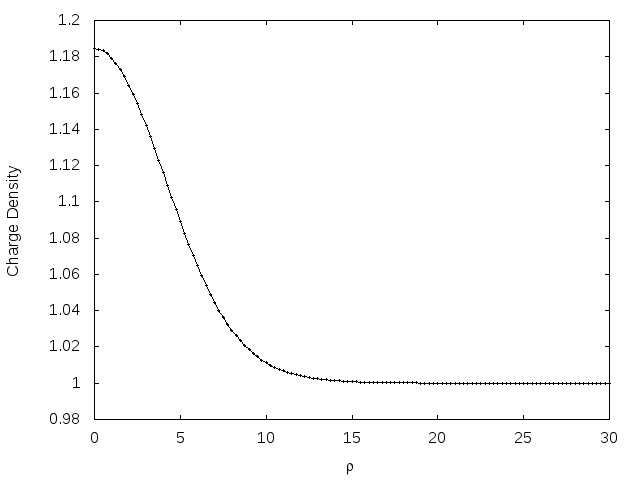}
\caption{The charge density ($\bar{\rho}$) as a function of the radius ($\rho$) with $\alpha^2 = 0.03$ and $B=0.1$ for an axially symmetric droplet. \label{Ec2Axbbb}}
\end{figure}

\begin{figure}[ht]
\includegraphics[width=0.49\textwidth]{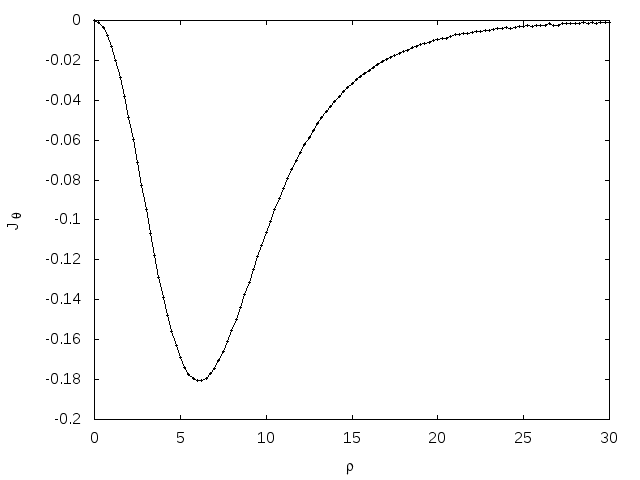}
\caption{The azimuthal current ($J_{\theta}$) as a function of the radius ($\rho$) with $\alpha^2 = 0.03$ and $B=0.1$ for an axially symmetric droplet. \label{EAxbbb}}
\end{figure}

 We can  examine several other quantities in the boundary theory, namely the charge density $\bar{\rho}$ and the azimuthal current $J_{\theta}$. Figure \ref{Ec2Axbbb} shows the charge density profile of the droplet. Figure \ref{EAxbbb} shows the axial current. We note that the formation of currents at the edge of the superconducting droplets is expected as the superconductor will attempt to expel any magnetic fields.

\clearpage

\section{Conclusion}

We have demonstrated the versatility of a gradient flow approach to holographic superconductors. In addition to reliably finding solutions to the equations of motion, we are also able to determine the stability of the solutions in a nonperturbative way. We find stable axially symmetric droplets with a constant background magnetic field, whereas the droplets found in \cite{albash} would enhance or weaken the magnetic field at the core depending on the temperature. The gradient flow is a much more general approach to finding droplets as we can consider any magnetic field as a fixed background specified by initial conditions. 

Although the flow parameter itself does not have a straightforward physical interpretation, we can exploit its connection with the energy to gain insight into how the system can undergo the phase change from hairless black hole (normal phase) to a black hole with scalar hair (superconducting phase) in a quasistatic way. \\

The AdS/CFT correspondence  provides an equivalence between the Euclidean on-shell action in the bulk and the free energy of the boundary. This means that at the fixed points of the flow the free energy of the bulk, $\E$, is equivalent to the free energy of the superconductor. Away from the fixed points the AdS/CFT dictionary does not tell us anything about the relationship between the two energies. However, using the gradient flow method, we can formally link static off-shell configurations in the bulk and in the boundary at the same value of the flow parameter $\tau$. For quasistatic evolution at least, it may be  reasonable to think of this link as an extension of the AdS/CFT correspondence. 
\\

The extension of the gradient flow method to include the metric flow would allow us to study the system away from the probe limit since the stationary points of the metric flow will be the fully backreacted metric. Although holographic superconductors with metric backreactions have been studied, it is unclear how allowing the metric to change will influence the stability of the scalar hair solutions. To study the system away from the probe limit it is necessary to include additional boundary terms in the action to regulate divergences of the free energy. The evolution of the metric from the vacuum black hole solution is a physically interesting problem in its own right, and this type of metric curvature flow with source terms is an interesting problem in mathematics.

Another application of gradient flow to AdS/CFT that is an active area of research is the Hamilton-Jacobi (HJ) first order equations \cite{HamiltonJacobi1,HamiltonJacobi2}. The HJ equations are a gradient flow generated by an on-shell action in the bulk using the AdS radial coordinate $r$ as the flow parameter. The solutions to the HJ equations are solutions to the equations of motion in the bulk for all $r$, however the flow itself is of interest due to its relationship with the renormalization group flow on the boundary. Similar to our approach, the HJ equations define a one parameter family of field configurations in the bulk; however since the flow parameter is the AdS radius each configuration represents a single timelike slice of AdS space at a given radius. Since the energy scale of the conformal field theory is effectively controlled by the AdS length scale, the slices of increasing radius correspond to increasing energy scales in the boundary theory. In this way the HJ gradient flow equations can be interpreted as the renormalization group flow on the boundary. This application differs from our approach in several ways, the main difference is that our gradient flow is generated by the off-shell action such that the on-shell configurations are fixed points of the flow. The solutions of the flow equations in our case are therefore not solutions to the dynamical, time-dependent equations of motion for the system. Recall that we are interested in approximating quasistatic evolution of thermal systems that are perturbed from equilibrium. In general such systems can be studied by solving the full dynamical equations of motion with appropriate boundary conditions allowing energy to flow through the black hole boundary. In the quasistatic case, where the system is never far from equilibrium, it may be possible to study the dynamics of the system using some approximation. The gradient flow method, which takes fields along the path of steepest descent towards the extremum of the action, is a good candidate.

{\bf ACKNOWLEDGMENTS}
This research was supported in part by the Natural Sciences and Engineering Research Council of Canada, and a University of Manitoba graduate fellowship.

\appendix
\section{Numerical Methods}\label{A1}

Since our gradient flow equations are analogous to the heat equation we solve them using a simple explicit finite difference scheme with forward differences in flow time and centered differences in our spatial coordinates. Such methods tend to be stable and convergent as long as the time step is proportionally smaller than the square of the space step,
\bea
d\tau \leq C du^2.\label{CFLcond}
\eea
For a simple one-dimensional heat equation it can be shown that $C=0.5$, but it is difficult to precisely compute the value of $C$ for the type of nonlinear coupled partial differential equations that we have solved.
We typically work with an equal number grid points in both $u$ and $x$ (or $\rho$). Since $u\in(0,1)$ and $x\in(0,30)$ we have $du < dx$ and therefore we do not need to consider a separate convergence condition of the form (\ref{CFLcond}) for $dx$.

We find that $d\tau$ obtained from \eqref{CFLcond} with $C=0.25$ is sufficient for good convergence.
Using this value of $C$ we see that the computer time required to reach $\tau=1$ is proportional to $du^{-3}$ for the one-dimensional case and $du^{-4}$ for two dimensions. To increase computational speed, we therefore want $du$ to be as large as possible without sacrificing accuracy.

\begin{figure}[]
\includegraphics[width=0.45\textwidth]{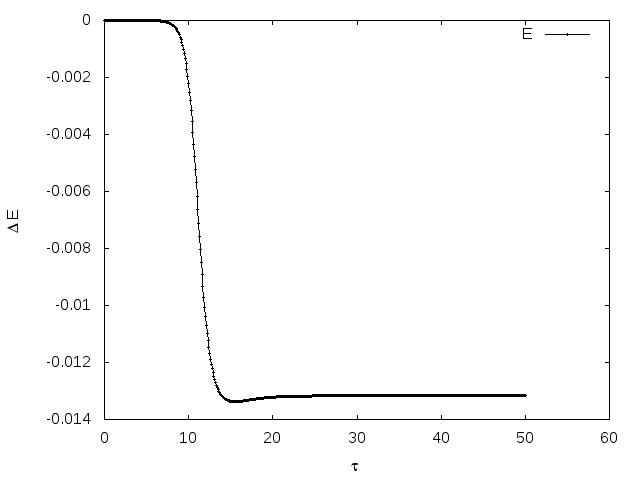}
\caption{$\Delta E$ as a function of $\tau$ with decreased grid spacing. \label{Etau300}}
\end{figure}

\begin{figure}[]
\includegraphics[width=0.45\textwidth]{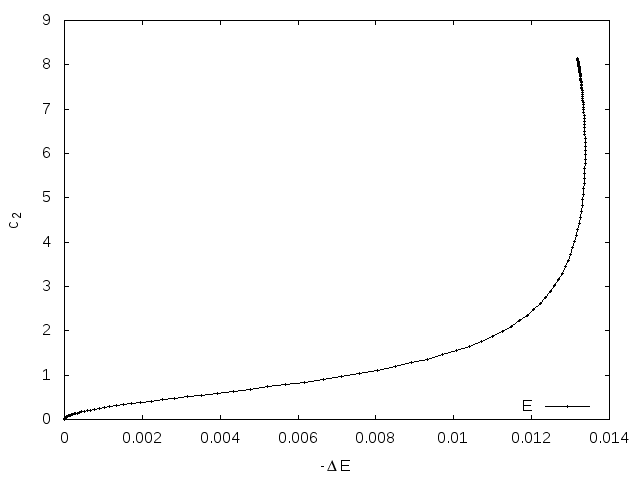}
\caption{The operator $c_2$ as a function of $\Delta E$. \label{Ec300}}
\end{figure}

Most of the results presented in this paper were calculated with a $300 \times 300$ grid. The exception is  Figs. \ref{vstau} and \ref{Ec2}. where we use 2000 grid points. In this case we found that with fewer grid points there was small but noticeable nonmonotonic behavior in $\Delta E$ which suggested that the endpoint of the flow did not minimize the free energy of the system. Figs. \ref{Etau300} and \ref{Ec300} show this behavior with 300 grid points. 
The nonmonotonicity is more noticeable at lower temperatures since $p(u)$ is larger in those cases, and the fact that it disappears when the grid spacing is reduced (see Figs. \ref{vstau} and \ref{Ec2}) proves that the effect is numerical. 
Although we do not see the same behavior in the spatially dependent cases we instead see smaller fluctuations of the energy, in particular when it is small relative to the grid spacing. These fluctuations are due to the larger error in the spatial derivatives since $dx>du$.

\begin{figure}[ht]
\begin{subfigure}[h]{0.49\textwidth}
\includegraphics[width=0.9\textwidth]{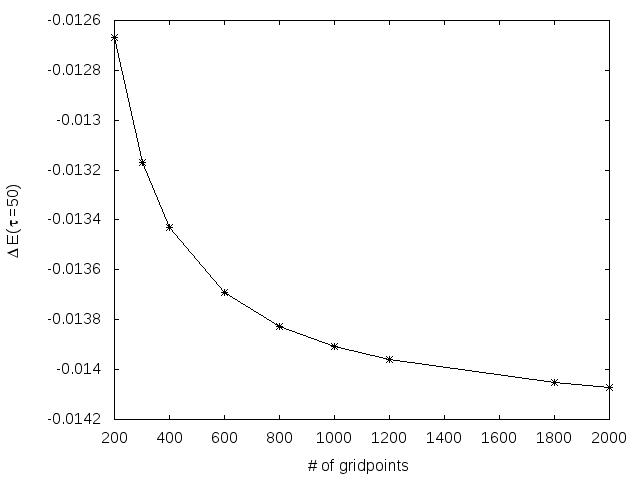}
\caption{$\Delta E$ as a function of the number of gridpoints for spatially independent fields \label{Econv}}
\end{subfigure}
\begin{subfigure}[h]{0.49\textwidth}
\includegraphics[width=0.9\textwidth]{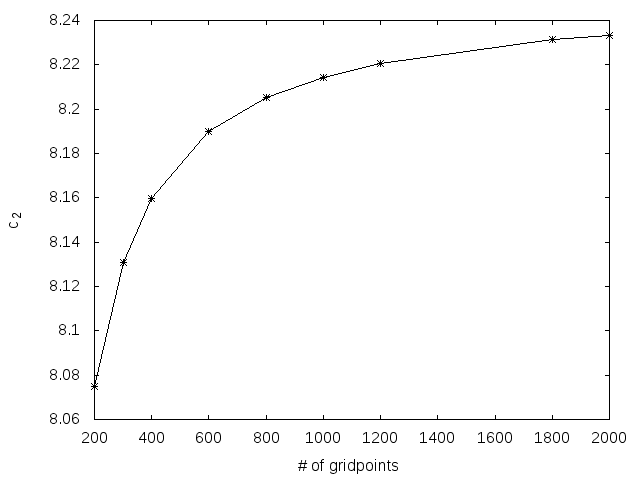}
\caption{$c_2$ as a function of the number of gridpoints \label{c2conv}}
\end{subfigure}
\begin{subfigure}[h]{0.49\textwidth}
\includegraphics[width=0.9\textwidth]{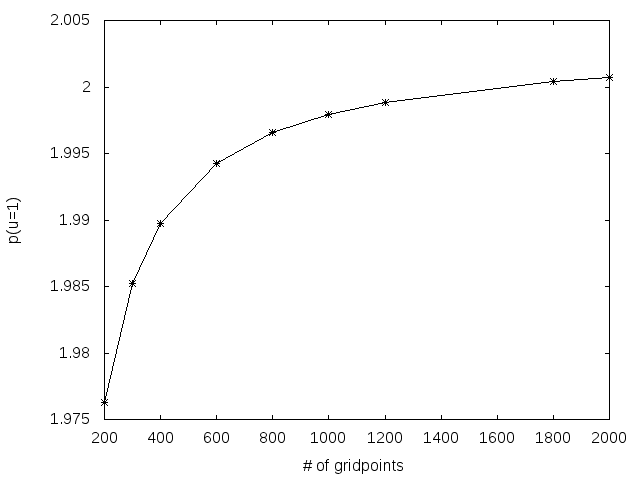}
\caption{$p(u=1)$ as a function of the number of gridpoints \label{pconv}}
\end{subfigure}
\begin{subfigure}[h]{0.49\textwidth}
\includegraphics[width=0.9\textwidth]{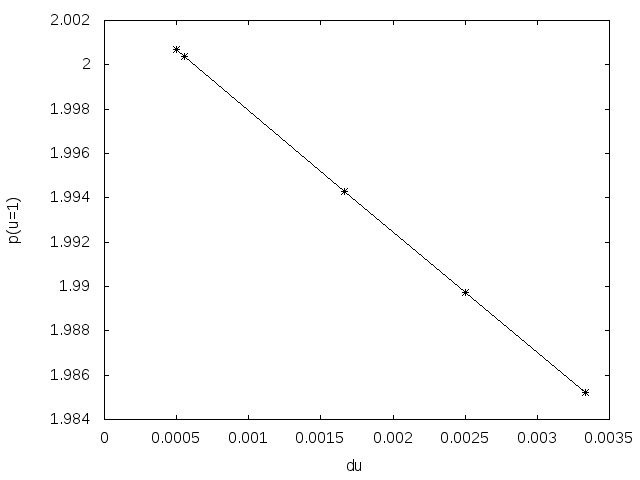}
\caption{$p(u=1)$ as a function of $du$ \label{pconvdu}}
\end{subfigure}
\caption{The dependence of several quantities on the number of grid points. \label{conv}}
\end{figure}

In Fig. \ref{conv} we look at the dependence of $c_2$, $\Delta E$ and $p(u=1)$ on the number of grid points used (equivalently $du^{-1}$), at the end of the flow for the spatially independent case where the computation time depends least on grid spacing. 
We see that quantities approach a fixed value as the number of grid points increases. We note that the range for the $y$ axis on these plots is roughly a $\% 10$ variation for the energy, and $\% 2$ for $c_2$ and $p(u=1)$. We expect the larger error in $\Delta E$ since it is obtained from a numerically computed integral of fields which introduces additional error. The fields themselves and $c_2$ depend mainly on the error in first order finite differences which is proportional to $du$. In Fig. 19(\subref{pconvdu}) we see that $p(u=1)$ is linear in $du$, and we can extrapolate to $du=0$ to find that $p(u=1)\approx 2.0034$ is within $\% 1$ of the value obtained using only $300$ grid points.\\

{\bf Boundary conditions}\\
Since we are using centered finite differences we need to take special care of the points at the boundaries. For all cases we treat the $u=0$ and $u=1$ boundaries in the same way.   
At the AdS boundary, $u=0$, all terms with derivatives are multiplied by a factor $u^2$. We assume that the derivatives of our fields are finite at $u=0$, which means that any term with a derivative does not contribute to the flow equations.
At the horizon, $u=1$, the factor $h(u)$ goes to 0. We  require that $A_t(u=1)=0$ in such a way that the ratio $\frac{A_t^2}{h(u)} =0$ at $u=1$. Any $u$ derivatives that we need are calculated using a one-sided finite difference. 

For the boundaries at $x_{\rm max}$ and $\rho_{\rm max}$ we can usually use one-sided finite differences since typically the fields are 
approximately constant at this boundary. The one exception is the field $A_{\theta}$ for which we enforce the condition $\partial_{\rho}A_{\theta}(\rho_{\rm max}) = B\rho$.  For a finite superconductor with radius $\rho_{\rm max}$, this is simply the condition that there is a fixed external magnetic field. 

The boundaries at $x=0$ and $\rho=0$ can be handled by adding an extra grid point at $x=-dx$ and calculating centered finite differences as usual. The value of the fields at this point is determined using the symmetry of the configuration. We take $A_t$ to be always symmetric, $A_y$ and $A_{\rho}$ are always antisymmetric, and $p$ is symmetric except for the soliton case, where it is antisymmetric.


\end{document}